\documentclass[a4paper,fleqn]{cas-dc}

\usepackage[numbers,sort&compress]{natbib}

\usepackage{bm}
\usepackage{comment}
\def\tsc#1{\csdef{#1}{\textsc{\lowercase{#1}}\xspace}}
\tsc{WGM}
\tsc{QE}
\tsc{EP}
\tsc{PMS}
\tsc{BEC}
\tsc{DE}

\usepackage{framed} % Framing content
\usepackage{multicol}
\usepackage{nomencl} % Nomenclature package
\makenomenclature
\renewcommand*\nompreamble{\begin{multicols}{2}}
\renewcommand*\nompostamble{\end{multicols}}

\usepackage{algorithm}
\usepackage{algpseudocode}
\usepackage{amsmath,amsfonts,amsthm}
\usepackage{graphicx}
\usepackage{subcaption}
\usepackage{url}
\usepackage{hyperref} % 使用 hyperref 包以支持 URL
\usepackage{xcolor}
\usepackage{makecell}
\usepackage{amsmath}
\usepackage{mhchem}
\hypersetup{
    colorlinks=true, % 将框变成颜色链接
    linkcolor=black, % 设置链接颜色
    filecolor=black, % 设置文件链接颜色
    urlcolor=black, % 设置 URL 链接颜色
    citecolor=black, % 设置引用颜色
    linkbordercolor={1 1 1}, % 无边框
    pdfborder={0 0 0} % 无边框
}

\begin{document}
\let\WriteBookmarks\relax
\def\floatpagepagefraction{1}
\def\textpagefraction{.001}
\let\printorcid\relax % 可去掉页面下方的ORCID(s)
\shorttitle{\rmfamily Energy-carbon comprehensive efficiency evaluation of hydrogen metallurgy system considering low-temperature waste heat recovery}
\shortauthors{\rmfamily Qiang~Ji et~al.}

% 题目
\title [mode = title]{Energy-carbon comprehensive efficiency evaluation of hydrogen metallurgy system considering low-temperature waste heat recovery}                      
% \tnotemark[1,2]
% Probabilistic Risk and 
% \tnotetext[1]{This document is the results of the research
%    project funded by the XXXX.}

% \tnotetext[2]{The second title footnote which is a longer text matter
%    to fill through the whole text width and overflow into
%    another line in the footnotes area of the first page.}

% 单位
\affiliation[1]{organization={State Key Laboratory of Power Power System Operation and Control, Department of Electrical Engineering, Tsinghua University},
                % addressline={Jawahar Nagar}, 
                city={Beijing},
%               citysep={}, % Uncomment if no comma needed between city and postcode
                postcode={100084}, 
                % state={Kerala},
                country={China}}

% \affiliation[2]{organization={Department of Earth and Environmental Engineering, Columbia University},
%                 % addressline={College Station}, 
%                 city={New York},
%                 postcode={10027}, 
%                 % state={Orissa}, 
%                 country={United States}}

\affiliation[2]{organization={School of Metallurgical and Ecological Engineering,  University of Science and Technology Beijing},
                % addressline={Street 29}, 
                postcode={100083}, 
                % postcodesep={}, 
                city={Beijing},
                country={China}}

% \affiliation[4]{organization={Department of Electrical and Computer Engineering, Texas A\&M University},
%                 addressline={College Station}, 
%                 city={TX},
%                 postcode={77843}, 
%                 % state={Orissa}, 
%                 country={United States}}

% 作者

% \author[1,3]{Yingrui Zhuang}[type=editor,
%                         auid=000,bioid=1,
%                         prefix=Sir,
%                         role=Researcher,
%                         orcid=0000-0001-0000-0000]
\author[1]{Qiang~Ji}[style=chinese]
\cormark[1]
\ead{jiq23@mails.tsinghua.edu.cn} 
\credit{Conceptualization, Methodology, Software, Writing}
\author[1]{Lin~Cheng}[style=chinese]
\credit{Conceptualization}
\author[2]{Zeng~Liang}[style=chinese]
\credit{Methodology, Software, Writing}
\author[1]{Yingrui~Zhuang}[style=chinese]
\credit{Reviewing, Editing, Writing}
\author[1]{Fashun~Shi}[style=chinese]
\credit{Editing, Supervision}
\author[2]{Jianliang~Zhang}[style=chinese]
\credit{Conceptualization}
\author[2]{Kejiang~Li}[style=chinese]
\credit{Supervision}

% \author[2]{Ning~Qi}[style=chinese]
% \credit{Methodology, Data Acquisition, Reviewing and Editing}

% \author[1]{Xinyi~Wang}[style=chinese]
% \credit{Data Curation, Software, and Writing}

% \author[3]{Yue~Chen}[style=chinese]
% \credit{Conceptualization, Methodology, Reviewing and Editing}

% \author[4]{Le~Xie}[style=chinese]
% \credit{Data curation, Writing - Original draft preparation}

\cortext[cor1]{Corresponding author}
% \cortext[cor2]{Principal corresponding author}
% \fntext[fn1]{This is the first author footnote, but is common to third
% \fntext[fn2]{Another author footnote, this is a very long footnote and
%   it should be a really long footnote. But this footnote is not yet
%   sufficiently long enough to make two lines of footnote text.}

% \nonumnote{
% This work was supported in part by 
% Special Foundation of Jiangsu Province Innovation Support Program (Soft Science Research) under Grant (No. BE2023093-1)
% and 
% National Natural Science Foundation of China under Grant (No. 52307144).
% }

%% 摘要
\begin{abstract}
To address the lack of energy-carbon efficiency evaluation and the underutilization of low-temperature waste heat in traditional direct reduction iron (DRI) production, this paper proposes a novel zero-carbon hydrogen metallurgy system that integrates the recovery and utilization of low-temperature and high-temperature waste heat, internal energy, and cold energy during hydrogen production, storage, reaction and circulation. Firstly, the detailed mathematical models are developed to describe energy and exergy characteristics of the operational components in the proposed zero-carbon hydrogen metallurgy system. Additionally, energy efficiency, exergy efficiency, and energy-carbon efficiency indices are introduced from a full life-cycle perspective of energy flow, avoiding the overlaps in energy inputs and outputs. Subsequently, the efficiency metrics of the proposed zero-carbon hydrogen metallurgy system are then compared with those of traditional DRI production systems with \ce{H2}/CO ratios of 6:4 and 8:2. The comparative results demonstrate the superiority and advancement of the proposed zero-carbon hydrogen metallurgy system. Finally, sensitivity analysis reveals that the overall electricity energy generated by incorporating the ORC and expander equipments exceeds the heat energy recovered from the furnace top gas, highlighting the energy potential of waste energy utilization.
\end{abstract}

% probabilistic power flow calculation method based on an improved Gaussian mixture model
% to analyze the risks of EVs (e.g., the low voltage).
% based on the probabilistic power flow calculation,
% for which we develop an improved Gaussian mixture model to enhance the calculating efficiency.

% \includegraphics{figs/cas-grabs.pdf}

\begin{highlights}
    \item A novel zero-carbon hydrogen metallurgy system with low-temperature waste heat recovery.
    \item The electricity energy generated by the ORC and expander exceeds the energy recovered from furnace top gas.
    \item Energy, exergy, and energy-carbon efficiencies evaluation for the zero-carbon hydrogen metallurgy system.
    % \item Formulation and corresponding convexification of real-time hosting capacity.
    %\item Energy-carbon efficiency evaluation method is proposed.
\end{highlights}


\maketitle

\renewcommand{\nomgroup}[1]{%
	\item[%\large
      \textbf{%
        \ifthenelse{\equal{#1}{A}}{\textit{Abbreviations}}{}%
        \ifthenelse{\equal{#1}{B}}{\textit{Sets}}{}%
        \ifthenelse{\equal{#1}{C}}{\textit{Parameters}}{}%
        % \ifthenelse{\equal{#1}{D}}{}{}%
        \ifthenelse{\equal{#1}{E}}{\textit{Functions}}{}%
	}]%
}
\nomenclature[C]{$W_i^{in},W_{i}^{out}$}{Input and output power of the component $\mathrm{i}$, respectively}
\nomenclature[C]{$EX_{i}^{in}, EX_{i}^{out}$}{Input and output exergy of the component $\mathrm{i}$, respectively}
\nomenclature[C]{$T_i^{in}$,$T_{i}^{out}$}{Inlet and outlet temperatures of the component $\mathrm{i}$, respectively}
\nomenclature[C]{$W_{reaction}$}{The heat absorbed by the reaction}
\nomenclature[B]{$N_{c}, N_{e}$}{Numbers of compression and expansion stages}
\nomenclature[B]{$\eta _c, \eta_{ex}$}{The isentropic efficiencies of compressor and expander, respectively}
\nomenclature[C]{$P_{comp,i}^{in}, P_{comp,i}^{out}$}{Inlet and output hydrogen pressures of the $\mathrm{i^{th}}$ stage compressor}
\nomenclature[C]{${M_{{H_2}}}$}{Molar mass of hydrogen}
\nomenclature[C]{$R$}{Ideal gas constant}
\nomenclature[C]{$h_i^{in}, h_i^{out}$}{Input and output enthalpies of the component $\mathrm{i}$, respectively}
\nomenclature[C]{$s_i^{in}, s_{i}^{out}$}{Input and output entropy of the component $\mathrm{i}$, respectively}
\nomenclature[C]{$m_i$}{The mass flow of substance $\mathrm{i}$ }
\nomenclature[C]{$P_{exp,i}^{in}, P_{exp,i}^{out}$}{Inlet and output hydrogen pressures of the $\mathrm{i^{th}}$ stage expander, respectively}
\nomenclature[C]{$\eta _{ex,i}$}{The expansion efficiency of the $\mathrm{i^{th}}$ stage expander}
\nomenclature[C]{$M_{Energy}$}{Energy baseline cost(CNY/J)}
\nomenclature[C]{$Q_c$}{Energy quantity(J)}
\nomenclature[C]{$C_{DRI}$}{Actual carbon emissions of DRI (t)}
\nomenclature[C]{$P_{pla}$}{Plasma heating power}
\nomenclature[C]{$Re_1$}{Plasma heating irreversible loss rate}
\nomenclature[C]{$W_i^{ph},W_{i}^{ch}$}{The physical and chemical energy of substance $\mathrm{i}$, respectively}
\nomenclature[C]{$EX_i^{ph}, EX_{i}^{ch}$}{The physical and chemical exergy of substance $\mathrm{i}$, respectively}
\nomenclature[C]{$n_1,n_2$}{The molar numbers of reduction and circle hydrogen, respectively}
\nomenclature[C]{$T_{ore}$}{Inlet temperature of pellet ore}
\nomenclature[C]{$w_{Fe},M_{Fe}$}{Mass fraction and molar mass of iron, respectively.}
\nomenclature[C]{$\Delta H$}{The enthalpy change of reduction reaction}
\nomenclature[C]{$\eta_1, \eta_2$}{Heat loss ratios from the shaft furnace and shaft dust, respectively}
\nomenclature[C]{$LHV_{\ce{H2}}$}{The lower heating value of hydrogen}
\nomenclature[C]{$n_{\ce{H2}}$}{Hydrogen production ratio of electrolyzer}
\nomenclature[C]{$n_c$}{Number of electrolyzer cells }
\nomenclature[C]{$V_{cell}$}{The voltage of the cell}
\nomenclature[A]{ORC}{Organic rankine cycle}
\nomenclature[C]{$\varphi_i$}{Molar fraction of the component $\mathrm{i}$ in the mass flow}
\nomenclature[C]{$Ex_i$}{Molar chemical exergy of the component}
\nomenclature[C]{$\eta_{H_{2}}$}{Utilization ratio of hydrogen}
\nomenclature[A]{$EC$}{Energy-carbon efficiency}
\nomenclature[C]{$\varphi_{CE}$}{Energy-carbon coupling factor}
\nomenclature[C]{$M_{\ce{CO2}}$}{Carbon emissions cost (CNY)}
\nomenclature[C]{$C_{base}$}{Carbon emissions allowance (t)}
\nomenclature[C]{$P_{\ce{CO2}}$}{Carbon trading price(CNY/t)}
\nomenclature[C]{$M_{c-p}$}{Energy price of production (CNY)}
\nomenclature[A]{CE}{Carbon emissions efficiency}
\nomenclature[C]{$\vartheta /\nu$}{Punishment factor}
\nomenclature[A]{DRI}{Direct reduction iron}
\nomenclature[C]{$\Delta G_i$}{Gibbs free energy of substance $\mathrm{i}$ }
\nomenclature[A]{HDRI}{Hot direct reduction iron}
\nomenclature[A]{EE}{Energy efficiency}
\nomenclature[A]{EXE}{Exergy efficiency}
\nomenclature[A]{EC}{Energy-carbon efficiency}
\nomenclature[A]{CET}{The equivalent energy of carbon emissions cost}
\begin{table*}[!t]
\begin{framed}
    {
        \fontfamily{ptm}\selectfont % 设置字体是新罗马
        \printnomenclature
    }
\end{framed}
\end{table*}
\section{Introduction}
\label{sec1}
As a backbone of national economic development, China's crude steel output reached 1032 Mt \cite{yu}, with 90\% produced using the blast furnace-basic oxygen furnace  (BF-BOF) process. The iron and steel industry contributed to 15\% of the total carbon emissions and 13.5\% of energy consumption \cite{lei}. With the development of carbon peak and carbon neutrality goals \cite{1}, China’s iron and steel industry \cite{2}, which heavily depends on the BF-BOF process, faces the challenges of significant carbon emissions and energy-intensive consumption \cite{3}. How to reduce carbon emissions and lower energy consumption has been becoming an urgent task for the iron and steel industry \cite{4}.\par
In terms of carbon reduction, the iron and steel industry with BF-BOF has explored strategies such as gas recycling, combined cooling, heating, and power (CCHP), and carbon capture, utilization, and storage (CCUS) \cite{5}. However, these measures only redirect the flow of carbon emissions without addressing the fundamental reliance on carbon-based reduction materials. Hydrogen, with its high calorific value, strong reducibility, and zero-carbon emissions, is regarded as the cleanest energy source. It can replace carbon-based reduction gases in DRI production \cite{6}. Currently, hydrogen-based DRI production technologies, such as MIDREX \cite{7} and Energiron-ZR \cite{9}, are widely used in regions with abundant and affordable natural gas, offering significant economic advantages \cite{10}. However, in China, where coal resources are abundant and natural gas is scarce, using natural gas as a feedstock for DRI production is not economically viable \cite{11}. Therefore, China’s DRI production industries, such as HBIS Zhangxuan and Baosteel Zhanjiang, utilize abundant coke oven gas as a feedstock to produce sponge iron \cite{12}. Although these industries reduce carbon emissions by utilizing coke oven as a feedstock, they remain heavily reliant on coal as their primary energy. From a full life-cycle perspective of carbon emissions, these measures fail to fundamentally address the issue of high carbon emissions, leaving their products less competitive compared to green steel. To effectively reduce the carbon footprint, many scholars \cite{13,14,15} have proposed using a mixed gas of green hydrogen and CO for DRI production. The green hydrogen produced from renewable energy resource is carbon-free, and the heat generated from the reaction between \ce{CO} and \ce{Fe2O3} can  maintain the shaft furnace's internal high-temperature. Although the mixture gas reduces carbon emissions and the raw material gas entering shaft furnace, high expenses of hydrogen production and carbon emissions cost are frequently criticized. Additionally, the statistics show that the installed capacity of renewable energy has exceeded 1.6 billion kilowatts in China \cite{configuration2024}, but its utilization ratio is below 10\%, leading to substantial wind and solar curtailment. Using surplus green electricity to produce hydrogen as a reduction gas offers an economically viable pathway to achieve a zero-carbon hydrogen metallurgy system.  \par
In terms of lowering energy consumption, efforts to improve energy efficiency have largely concentrated on the individual equipment and process. For example, \cite{16} uses nitrogen as a heat carrier due to its higher heat capacity, providing heat to shaft furnace to maintain the high temperature. After a complex reduction and combustion process, \cite{add16} recovers waste heat of the furnace top gas to reduce energy consumption of the shaft furnace. \cite{xu} investigates various ratios of \ce{H2}/CO to optimize energy utilization of the reduction gas. \cite{bai} integrates an oxygen blowing device into the shaft furnace, where the exothermic reaction significantly contributes to heat supply, enhancing hydrogen utilization and overall energy efficiency. Although these measures certainly improve energy efficiency of shaft furnace, little attention has been paid on the whole DRI production process and the low-temperature waste heat recovery \cite{add1}. However, waste heat of production process of the iron and steel industry accounts for 30\% of the total energy consumption \cite{zhang}. Therefore, incorporating low-temperature waste heat recovery into the overall zero-carbon hydrogen metallurgy system is essential for reducing energy consumption and improving energy efficiency. \par
Energy and exergy efficiencies are widely used to evaluate energy utilization characteristics in the DRI production process. Many existing publications evaluating energy efficiency often rely on prediction methods such as data envelopment analysis (DEA), slack-based measure, and DEA extensions \cite{19,20,21}. These approaches use regression techniques to construct an “efficient frontier” and predict the efficiency of various decision-making units. However, these measures frequently overlook potential interactions among input variables and intermediate variables, and the DRI production process is a complex, multi-energy flow coupling system. This limitation leads to deviation in the energy efficiency evaluation. Additionally, these methods function as black-box models, which are better suited for planning and design phases but less effective for guiding existing system in selecting strategies to improve energy efficiency. Some studies have assessd the energy and exergy efficiencies by aggregating the inputs and outputs of individual components in the DRI  production process \cite{22,281,23}. However, this aggregation way often neglects full life-cycle of energy flow transferring, transporting and circulating, leading to overlaps that obscure the distinction of physical and chemical energies. Consequently, the obtained energy efficiency may depart from the actual system performance \cite{24}. Furthermore, with the development of China’s carbon market, carbon trading prices is becoming an additional constraint on the profitability of the iron and steel industry \cite{25,26}. While some researchers \cite{27,28,29} focus solely on carbon efficiency as a criterion to evaluate carbon emissions, overlooking the economic effect of carbon emissions. Few studies have integrated energy efficiency with carbon efficiency by economic properties as an index to evaluate the system’s energy utilization characteristics. This gap limits the ability of smaller iron and steel enterprises to identify the most effective investment strategies for improving energy-carbon efficiency, hindering their transition toward low-carbon and high-quality development. \par
Motivated by these research gaps, this paper proposes hydrogen as a bridge linking the raw material consumption of the iron and steel industry and the efficient absorption of renewable energy. By integrating low-temperature waste heat recovery, the zero-carbon hydrogen metallurgy system is introduced, along with comprehensive evaluation indexes in energy efficiency, exergy efficiency, energy-carbon efficiency. The main contributions of this study are as follows:\par
1) An innovative and comprehensive zero-carbon hydrogen metallurgy system is proposed. Beyond conventional DRI production system units and equipment, we innovatively incorporate ORC process as well as high-pressure hydrogen expander process into the hydrogen metallurgy system. To the best of the author's knowledge, no such compressive research has been conducted in the literature. \par
2) This paper presents detailed models to describe energy and exergy characteristics of each component in the proposed zero-carbon hydrogen metallurgy system. From a full life-circle perspective of energy flow, energy and exergy efficiencies evaluations are conducted, ensuring no overlap in energy inputs and outputs.\par
3) By incorporating carbon emissions cost and penalizing the system's energy output, we propose a energy-carbon efficiency evaluation method. Energy, exergy, and energy-carbon efficiencies of the proposed zero-carbon hydrogen metallurgy system are compared with those of the traditional DRI production process with \ce{H2} /CO ratios of 6:4 and 8:2, highlighting superiority and advancement of the former. \par
The rest paper is organized as follows: Section 2 provides a detailed description of the proposed hydrogen metallurgy system with its energy-material balance. The detailed energy and exergy models of each component are defined in Section 3, along with the metrics for energy efficiency, exergy efficiency, and energy-carbon efficiency. The comparative studies and sensitivity analyses are carried out in Section 4. Conclusions are drawn in Section 5.
%% Labels are used to cross-reference an item using \ref command.
\section{Zero-carbon hydrogen metallurgy production system}
\label{sec2}
%% Use \subsection commands to start a subsection.
%\subsection{Example Subsection}
%\label{subsec1}
The zero-carbon hydrogen metallurgy system is shown in Fig.  1. It comprises renewable energy sources, alkaline water electrolyzers, compressors and  expanders, hydrogen storage tanks, an ORC unit, a plasma heater, a shaft furnace, low-and high- temperature thermal storage tanks, dust removal equipment, steam removal equipment, and other auxiliary components. During the reduction hydrogen producing and storage phase, the alkaline electrolyzers utilize green electricity to produce hot hydrogen at 98℃, which is firstly heat exchanged to produce hydrogen at ambient temperature, then compressed by three-stage compressor (the compression ratio of compressor is limited under 5 to ensure lower energy consumption) and stored in a high-pressure hydrogen tank. Throughout this process, both the heat from the hot hydrogen and the heat generated by every stage of three-stage compressor are captured and stored in a low-temperature thermal storage tank. In the reaction phase between high-temperature hydrogen and pellet ore, hydrogen is firstly released from the high-pressure storage tank, and the associated pressure drop drive three-stage expander to generate electricity. The released cold hydrogen facilitates the condensation and liquefaction of the working fluid in the ORC system, which simultaneously generates electricity by utilizing the low-temperature waste heat and ambient air to convert the working fluid from liquid to vapor. Next, a substantial amount of ambient temperature hydrogen is heated by a plasma heater and fed into the shaft furnace to produce hot direct reduction iron (HDRI). In the furnace top gas recycling phase, the high-temperature thermal storage tank recovers heat from furnace top gas, which do not participate in the reduction process but is used to supply heat. Subsequently, both dust removal and steam removal are employed to filter out dust and water vapor in the furnace top gas. Finally, the compressor stores the recycled hydrogen back into the high-pressure storage tank. Hydrogen recycling minimizes energy loss and extends the operational lifespan of the shaft furnace and refractory materials. It can be seen that hydrogen serves two key functions: it acts as a reducing gas, extracting oxygen molecules from pellet ore, and as a heat carrier, providing the necessary heat to maintain the internal temperature of shaft furnace. When the internal temperature of the shaft furnace exceeds 570°C, the reduction hydrogen undergoes a three-stage \cite{31} process as it rises from the medium to the top of the shaft furnace. The detailed reactions are as follows:

\begin{figure*}[t!]
\centering
\includegraphics[width=0.85\textwidth]{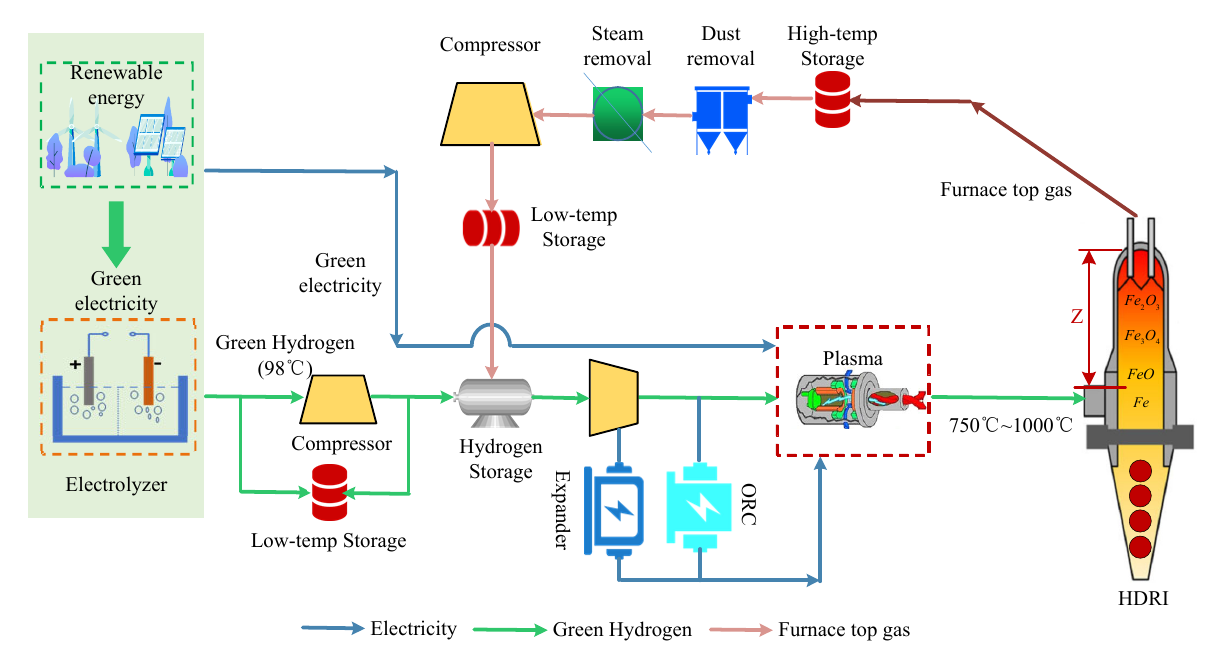}
\setlength{\abovecaptionskip}{-0.1cm}  
\setlength{\belowcaptionskip}{-0.1cm} 
\caption{The zero-carbon hydrogen metallurgy production system}
\captionsetup{justification=centering}
\vspace{-0.6cm}
\end{figure*}

\noindent \textbf{first stage:} \ce{Fe2O3 ->Fe3O4} \par
\begin{equation}
\begin{split}
 \frac{1}{2}\mathrm{Fe}_{2} \mathrm{O}_{3}+(n-\frac{4}{3}) \mathrm{H}_{2}(\ce{CO})= \frac{1}{3} \ce{Fe3O4} +\frac{1}{6} \mathrm{H}_{2} \mathrm{O}(\ce{CO2}) \\
 +(n-\frac{3}{2})\ce{H_2}(\ce{CO})
\end{split}
\end{equation}
\noindent \textbf{second stage:} \ce{Fe3O4 -> FeO} 
\begin{equation}
\begin{split}
\frac{1}{3}\ce{Fe3O4}+(n-1)\ce{H2}(\ce{CO})= \ce{FeO}+
\frac{4}{3}\ce{H2O}(\ce{CO2})\\
+(n-\frac{4}{3})\ce{H2}(\ce{CO})
\end{split}
\end{equation}
\noindent \textbf{third stage:} \ce{FeO -> Fe}
\begin{equation} 
 \ce{FeO}+n\ce{H2}(\ce{CO})= \ce{Fe}+\ce{H2O}(\ce{CO2})+(n-1)\ce{H2}(\ce{CO})
\end{equation}
\noindent \textbf{total reaction:}
\begin{equation}
\frac{1}{2}\ce{Fe2O3}+\frac{2}{3}\ce{H2}= \ce{Fe}+\frac{3}{2}\ce{H2O}(\ce{CO2})
\end{equation} \par
To avoid the risk of pellet ore sticking or melting, as well as DRI quality due to excessively high-temperature hydrogen, this study sets the hydrogen temperature range at 750°C to 1000°C.The heat sources within the shaft furnace consist of hydrogen and pellet ore. The heat outputs of shaft furnace include the heat consumed in the reduction reaction, the heat carried by the HDRI, furnace top gas, heat losses, shaft dust, and water vapor. The specific heat balance is detailed as follows:
%equation5-15;
\begin{equation}
W_{sf,toll}^{in}=W_{sf,H_{2}}^{in}+W_{sf,ore}^{in}
\end{equation}
\begin{equation}
    \begin{split}
    W_{sf,toll}^{out}=W_{reaction}+W_{DRI}^{out} +W_{CH_{2}}^{out}
    +W_{sf,loss} \\
    +W_{dust}^{out}+W_{H_{2}O}^{out}
    \end{split}
\end{equation}
\begin{equation}
    W_{sf,H_{2}}^{in}=(n_{1}+n_{2})\cdot c_{p,H_{2}}\cdot(T_{sf}^{in}-T_{0})
\end{equation}
\begin{equation}
W_{sf,ore}^{in}=m_{ore}\cdot
c_{ore}\cdot(T_{ore}^{in}-T_{0})
\end{equation}
\begin{equation}
W_{reaction}=\eta_{Fe-H_{2}}\cdot\frac{w_{Fe}}{M_{Fe}}\cdot m_{DRI}\cdot\Delta H
\end{equation}
\begin{subequations}
    \begin{align}
    \Delta H=H(Fe+\frac{3}{2}H_{2}O)-H(\frac12Fe_{2}O_{3}+\frac32H_{2})\\
    \Delta H=H(Fe+\frac{3}{2}\ce{CO2})-H(\frac12Fe_{2}O_{3}+\frac32\ce{CO})
    \end{align}
\end{subequations}
\begin{equation}
W_{DRI}^{out}=m_{DRI}\cdot c_{DRI}\cdot(T_{DRI}-T_{0})
\end{equation}
\begin{equation}
W_{CH_{2}}^{out}=n_{2}\cdot c_{p,H_{2}}\cdot(T_{sf}^{out}-T_{0})
\end{equation}
\begin{equation}
W_{sf,loss}=\eta_{1}\cdot W_{toll}^{out}
\end{equation}
\begin{equation}
W_{dust}^{out}=\eta_{2}\cdot m_{dust}\cdot c_{p,dust}\cdot(T_{sf}^{out}-T_{0})
\end{equation}
\begin{equation}
W_{H_{2}O}^{out}=\frac{\gamma_{H_{2}O}}{\gamma_{H_{2}O}-1}\cdot\frac{n_{1}}{2}\cdot R\cdot(T_{sf}^{out}-T_{0})
\end{equation} \par
In the aforementioned researches, the furnace top gas temperature and direct reduction iron temperature were typically treated as constant. However, this paper utilizes gas-sold differential equations \cite{Shao} to simulate and iteratively calculate the temperatures of furnace top gas and DRI. The detailed models are as follows. The iterative process of gas-solid heat transfer is illustrated in Fig. 2, while Fig. 3 presents the furnace top gas temperatures and HDRI temperatures across a reduction gas temperature range at 750°C to 1000°C.
\begin{equation}
    \frac{{d{T_{H_{2}}}}}{{dz}} = \frac{{6(1 - \varepsilon )}}{{dp}}\frac{S}{{{\rho _{H_{2}}}G}}\frac{{{h_p}}}{{{c_{H_{2}}}}}(T_{H_{2}} - {T_{DRI}}) + \frac{{\pi D}}{{{\rho _{H_{2}}}G}}\frac{{{h_w}}}{{{c_{H_{2}}}}}({T_{H_{2}} - {T_0}})
\end{equation}
\begin{equation}
    \frac{{d{T_{DRI}}}}{{dz}} = \frac{{6(1 - \varepsilon )}}{{dp}}\frac{S}{W}\frac{{{h_p}}}{{{c_{DRI}}}}({T_{H_{2}} - {T_{DRI}}})
\end{equation}
\begin{equation}
    \frac{{{h_p}{d_p}}}{{{k_{H_{2}}}}} = 2 + 0.39{\mathop{\rm Re}\nolimits} _p^{0.5}{\textit{Pr} ^{0.33}}
\end{equation}
\begin{equation}
    {{\mathop{\rm Re}\nolimits} _p} = \frac{{{\rho _{H_{2}}}{u_{H_{2}}}{d_p}}}{{{\mu _{H_{2}}}}}
\end{equation}
\begin{equation}
    \Pr  = \frac{c_{H_{2}}{\mu _{H_{2}}}}{k_{H_{2}}}
\end{equation}
Where $T_{H_{2}}$ is the bulk temperature of the hydrogen and $z$ is the vertical downward direction, $\varepsilon$, $dp$, $S$, $G$, $\rho_{H_{2}}$, $C_{g}$ and {D} are the bed porosity, pellet diameter, cross-sectional area of reduction zone, volumetric flow rate of hydrogen, density of hydrogen, specific heat capacity of hydrogen and inner diameter of reduction zone, respectively. The variables $h_{p}$ and $h_{w}$ are the gas-solid and overall heat transfer coefficients through the furnace wall, while $T_{DRI}$ and $T_{0}$ are the DRI and ambient temperatures. $W$ and $c_{DRI}$ are the mass flow rate and specific heat capacity of DRI. $k_{H_{2}}$, $Re_{p}$, $Pr$, $\mu_{H_{2}}$ $u_{H_{2}}$ are the coefficient of hydrogen thermal conductivity, Reynolds number, Prandtl number, the viscosity of hydrogen, superficial velocity, respectively.   
\begin{figure}[ht]
\centering
\includegraphics[width=0.45\textwidth]{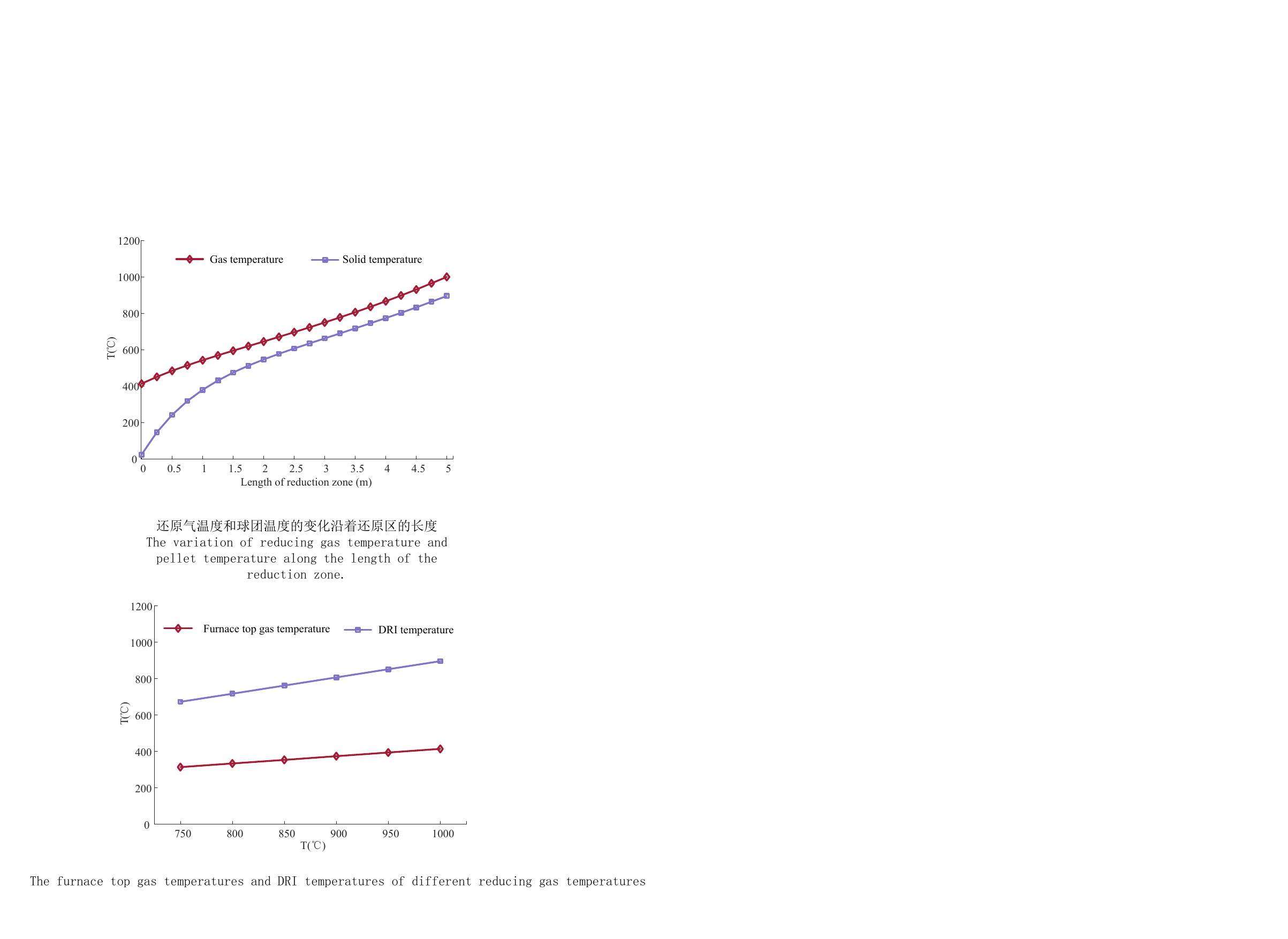}
\setlength{\abovecaptionskip}{-0.1cm}  
\setlength{\belowcaptionskip}{-0.1cm} 
\caption{The variations of reducing gas temperature and pellet temperature along the length of the reduction zone}
\captionsetup{justification=centering}
\vspace{-0.4cm}
\end{figure}
\begin{figure}[ht]
\centering
\includegraphics[width=0.45\textwidth]{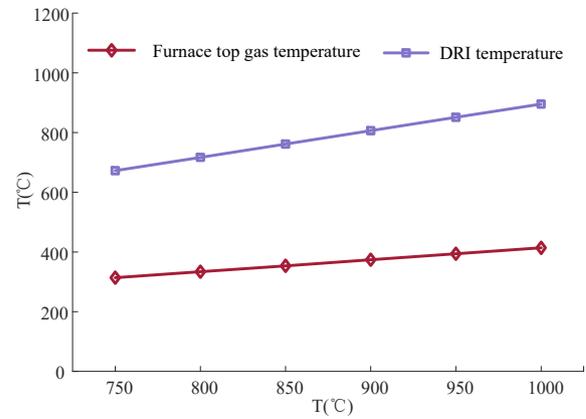}
\setlength{\abovecaptionskip}{-0.1cm}  
\setlength{\belowcaptionskip}{-0.1cm} 
\caption{The furnace top gas temperatures and DRI temperatures of different reducing gas temperatures}
\captionsetup{justification=centering}
\vspace{-0.4cm}
\end{figure}
\par The assumptions for the proposed zero-carbon hydrogen metallurgy system are as follows: all the physical and chemical reaction processes are in equilibrium states; the hydrogen storage tank follows the ideal gas law, with compression and expansion processes assumed to be isentropic and adiabatic; the reaction pressure within the shaft furnace is set to 8 bar, and reactions between reduction hydrogen and other ore components are disregarded; all furnace dust is assumed to be carried with the furnace top gas; the energy consumption of the dust removal and water removal devices is considered negligible. 
\section{Research mathematical model}
\label{sec3}
\subsection{Energy and exergy description of components}
\noindent 1) Alkaline water electrolyzer \par
When a direct current is charged to an alkaline water electrolyzer, hydrogen is produced at the cathode and oxygen at the anode, respectively \cite{32,33}. The chemical equation for this reaction is as follows:
%%electrolyzer
\begin{equation}
\text { Cathode : } 2 \mathrm{H}_{2} \mathrm{O}+2 e^{-} \rightarrow \mathrm{H}_{2} \uparrow+2 \mathrm{OH}^{-}
\end{equation}
\vspace{-0.5cm}
\begin{equation}
\text { Anode: } 2 \mathrm{OH}^{-} \rightarrow \mathrm{H}_{2} \mathrm{O}+\frac{1}{2} \mathrm{O}_{2} \uparrow+2 e^{-}
\end{equation}
\vspace{-0.5cm}
\begin{equation}
\text{Total reaction: }\mathrm{H}_{2} \mathrm{O} \rightarrow \mathrm{H}_{2} \uparrow+\frac{1}{2} \mathrm{O}_{2} \uparrow
\end{equation}
\par
The water electrolyzer is an endothermic process; however, due to ohmic and activation resistance, the cell voltage  is typically higher than the thermoneutral electrolysis voltage. Consequently, excess energy is consumed and dissipated during the reaction, leading to inevitable energy losses. The detailed energy inputs and outputs of the electrolyzer are as follows:
\begin{equation}
W_{\text {el }}^{\text {in }}=n_c\cdot V_{cell}\cdot I\cdot \Delta t
\end{equation}
\begin{equation}
W_{\text {el }}^{\text {out }}=L H V_{H_{2}} \cdot n_{H_{2}}+n_{H_{2}} \cdot c_{p, H_{2}} \cdot T_{H_{2}}^{\text {out }}
\end{equation}\par
The exergy input and output of the electrolyzer are electricity energy and hot hydrogen, respectively.
\begin{equation}
E X_{\text {el }}^{\text {in }}=W_{\text {el }}^{\text {in }}
\end{equation}
\begin{align}
 E X_{e l}^{\text {out }}=L H V_{H_{2}}\cdot n_{H_{2}}+n_{H_{2}} \cdot c_ {p,\ce{H2}} \cdot(T_{H_{2}}^{\text {out }}-T_{0}-\nonumber \\ &\hspace{-3cm} T_{0}\ln(T_{H_{2}}^{\text {out }}/T_{0}))
\end{align}
2) Compressor\par
Based on the isentropic efficiency \cite{34} \cite{35}, along with the constraints of the  compressor ratio and energy consumption, the input power and outlet hydrogen temperature of the  \textit{i}th stage compressor are as follows:
\begin{align}
\vspace{0.1cm} W_{\exp , i}^{i n}=\frac{\gamma}{\gamma-1} n_{\exp } R T_{\exp i}^{i n}\left(1-\left(\frac{P_{\exp , i}^{\text {out }}}{P_{\exp , i}^{i n}}\right)^{\frac{\gamma-1}{\gamma}}\right) \eta_{\operatorname{ex}, i}
\end{align}
\begin{equation}
T_{comp,i}^{out} = T_{comp,i}^{in}(1 + {{({{(\frac{{P_{comp,i}^{out}}}{{P_{comp,i}^{in}}})}^{\frac{{\gamma  - 1}}{\gamma }}} - 1)} \mathord{\left/
 {\vphantom {{({{(\frac{{P_{comp,i}^{out}}}{{P_{comp,i}^{in}}})}^{\frac{{\gamma  - 1}}{\gamma }}} - 1)} {{\eta _c}}}} \right.
 \kern-\nulldelimiterspace} {{\eta _c}}})
\end{equation}
\begin{equation}
W_{comp}^{in} = \sum\limits_{i = 1}^{{N_c}} {W_{comp,i}^{in}} 
\end{equation}
\begin{equation}
W_{comp}^{out} = W_{comp}^{in}
\end{equation}\par
The exergy entering the compressor is consumed by electricity, while the its output exergy is carried away by the hydrogen. The exergy input and output of the \textit{i}th stage compressor are as follows :
\begin{equation}
EX_{comp,i}^{in} = W_{comp,i}^{in}
\end{equation}
\begin{align}
EX_{comp,i}^{out} = {n_c}\frac{\gamma }{{\gamma  - 1}}R(T_{comp,i}^{out} - T_{comp,i}^{in})- \nonumber \\
&\hspace{-6cm}{T_0}({n_c}\frac{\gamma }{{\gamma  - 1}}R\ln \frac{{T_{comp,i}^{out}}}{{T_{comp,i}^{in}}} - {n_c}R\ln \frac{{P_{comp,i}^{out}}}{{P_{comp,i}^{in}}})
\end{align}
\begin{equation}
EX_{comp}^{in} = \sum\limits_{i = 1}^{{N_c}} {EX_{comp,i}^{in}}
\end{equation}
\begin{equation}
EX_{comp}^{out} = \sum\limits_{i = 1}^{{N_c}} {EX_{comp,i}^{out}}
\end{equation}
3) Thermal storage tank \par
Heat exchangers complete the recovery and release of thermal energy between hydrogen and working fluid. In the recovery stage of thermal storage tank, the energy input and output are as follows:
\begin{equation}
W_{C,TS}^{in} = \frac{\gamma }{{\gamma  - 1}}{n_c}R(T_{c,H2}^{in} - T_{c,H2}^{out})
\end{equation}
\begin{equation}
W_{C,TS}^{out} = {m_{c,f}}{c_f}(T_{c,f}^{out} - T_{c,f}^{in})
\end{equation}
the exergy input and output are:
\begin{align}
EX_{C,TS}^{in} = \frac{\gamma }{{\gamma  - 1}}{n_c}R(T_{c,H2}^{in} - T_{c,H2}^{out})- \nonumber \\
&\hspace{-4cm}- \frac{\gamma }{{\gamma  - 1}}{n_c}R{T_0}\ln (T_{c,H2}^{in}/T_{c,H2}^{out})
\end{align}
\begin{equation}
EX_{C,TS}^{out} = {m_{c,f}}(h_{c,f}^{out} - h_{c,f}^{in}) - {m_{c,f}}{T_0}(s_{c,f}^{out} - s_{c,f}^{in})
\end{equation}
\begin{equation}
h_{c,f}^{out} - h_{c,f}^{in} = {c_f}(T_{c,f}^{out} - T_{c,f}^{in})
\end{equation}
\begin{equation}
s_{c,f}^{out} - s_{c,f}^{in} = {c_f}\ln (T_{c,f}^{out}/T_{c,f}^{in})
\end{equation}
On the contrary, in the release stage of thermal storage tank, the energy input and output are as follows:
\begin{equation}
W_{D,TS}^{in} = {m_{d,f}}{c_f}(T_{d,f}^{in} - T_{d,f}^{out})
\end{equation}
\begin{equation}
W_{D,TS}^{out} = \frac{\gamma }{{\gamma  - 1}}{n_c}R(T_{d,H2}^{out} - T_{d,H2}^{in})
\end{equation}
the exergy input and output are:
\begin{equation}
EX_{D,TS}^{in} = {m_{d,f}}(h_{d,f}^{in} - h_{d,f}^{out}) - {m_{d,f}}{T_0}(s_{d,f}^{in} - s_{d,f}^{out})
\end{equation}
\begin{equation}
h_{d,f}^{in} - h_{d,f}^{out} = {c_f}(T_{d,f}^{in} - T_{d,f}^{out})
\end{equation}
\begin{equation}
s_{d,f}^{in} - s_{d,f}^{out} = {c_f}\ln (T_{d,f}^{in}/T_{d,f}^{out})
\end{equation}
\begin{align}
EX_{D,TS}^{out} = \frac{\gamma }{{\gamma  - 1}}{n_c}R(T_{d,H2}^{out} - T_{d,H2}^{in}) - \nonumber \\
&\hspace{-3cm} \frac{\gamma }{{\gamma  - 1}}{n_c}R{T_0}\ln (T_{d,H2}^{out}/T_{d,H2}^{in})
\end{align}
4) Expander \par
Based on the isentropic efficiency and the constraints of the expander ratio, the input potential power and outlet hydrogen temperature of the \textit{i}th stage expander are as follows:
\begin{equation}
W_{\exp ,i}^{in} = \frac{\gamma }{{\gamma  - 1}}{n_{\exp }}RT_{\exp ,i}^{in}(1 - {(\frac{{P_{\exp ,i}^{in}}}{{P_{\exp ,i}^{out}}})^{\frac{{1 - \gamma }}{\gamma }}}) \times {\eta _{ex,i}}
\end{equation}
\begin{equation}
T_{\exp ,i}^{out} = T_{\exp ,i}^{in}(1 - (1 - {(\frac{{P_{\exp ,i}^{in}}}{{P_{\exp ,i}^{out}}})^{\frac{{1 - \gamma }}{\gamma }}}) \times {\eta _{ex,i}})  
\end{equation}
\begin{equation}
W_{\exp }^{in} = \sum\limits_{i = 1}^{{N_e}} {W_{\exp ,i}^{in}}
\end{equation}
\begin{equation}
W_{\exp }^{in} = W_{\exp }^{out}
\end{equation}
the exergy input and output are:
\begin{align}
\hspace{-1.0cm} EX_{\exp ,i}^{in} = \frac{\gamma }{{\gamma  - 1}}{n_{\exp }}R(T_{\exp ,i}^{in} - T_{\exp ,i}^{out})- \nonumber \\
&\hspace{-4.0cm} {T_0}(\frac{\gamma }{{\gamma  - 1}}{n_{\exp }}R\ln \frac{{T_{\exp ,i}^{in}}}{{T_{\exp ,i}^{out}}} - {n_{\exp }} \cdot R\ln \frac{{P_{\exp ,i}^{in}}}{{P_{\exp ,i}^{out}}})
\end{align}
\begin{equation}
EX_{\exp ,i}^{out} = W_{\exp ,i}^{in}
\end{equation}
\begin{equation}
EX_{\exp }^{in} = \sum\limits_{i = 1}^{{N_e}} {EX_{\exp ,i}^{in}}
\end{equation}
\begin{equation}
EX_{\exp }^{out} = \sum\limits_{i = 1}^{{N_e}} {EX_{\exp ,i}^{out}}
\end{equation}
5) The ORC unit \cite{36}\par
The operating principle of the ORC is as follows: the liquid working fluid absorbs heat from ambient air and the low-temperature thermal storage tank, causing it to vaporize \cite{37}. The working vapor then enters into the expander, where it drives a generation to produce electricity. The expanded gas flows into the condenser, where it is cooled by low-temperature hydrogen, condensing the working fluid back into a liquid. Finally, the liquid working fluid is pumped back to the evaporator. The detailed process is shown in the Fig. 4. Additionally, the energy consumption in the condenser and pumping processes is neglected.\\
\textbf{Evaporator}, from 4 to 1 in the Fig. 4:
\begin{equation}
W_{orc,eva}^{in} = \sum\limits_{i \in (TS,Air)} {{m_{i,r14}}{c_{i,14}}(T_{i,ORC}^{s,in} - T_{i,ORC}^{s,out})}
\end{equation}
\begin{align}
\hspace{-0.8cm}W_{orc,eva}^{out} = {m_{r32}}({h_1} - {h_4}) = {m_{r32}}{c_{r32}}(T_{r32,eva}^{out} - T_{r32,eva}^{in})
\end{align}
\begin{equation}
W_{orc,eva}^{in} = W_{orc,eva}^{out}
\end{equation}
\begin{equation}
EX_{orc,eva}^{in} = W_{orc,eva}^{in}
\end{equation}
\begin{align}
\hspace{-1.0cm} EX_{orc,eva}^{out} = {m_{r32}}{c_{r32}}(T_{r32,eva}^{out} - T_{r32,eva}^{in}) - \nonumber \\
&\hspace{-3.2cm} {m_{r32}}{c_{r32}}T_{r32,eva}^{in}\ln( T_{r32,eva}^{out}/T_{r32,eva}^{in})
\end{align}
\textbf{Expander}, from 1 to 2 in the Fig. 4:
\begin{align}
\hspace{-0.8cm} W_{orc,\exp }^{in} = {m_{r32}}({h_1} - {h_2}) = {m_{r32}}{c_{r32}}(T_{r32,\exp }^{in} - T_{r32,\exp }^{out})
\end{align}
\begin{equation}
W_{orc,\exp }^{out} = W_{orc,\exp }^{in}
\end{equation}
\begin{equation}
EX_{orc,\exp }^{in} = W_{orc,\exp }^{in}
\end{equation}
\begin{align}
\hspace{-1.0cm} EX_{orc,\exp }^{out} = {m_{r32}}{c_{r32}}(T_{r32,\exp }^{in} - T_{r32,\exp }^{out})- \nonumber \\
&\hspace{-3.2cm} {m_{r32}}{c_{r32}}T_{r32,\exp }^{out}\ln (T_{r32,\exp }^{in}/T_{r32,\exp }^{out})
\end{align} \par
The overall energy and exergy inputs and outputs of the ORC unit are :
\begin{equation}
W_{orc}^{in} = W_{orc,eva}^{in}
\end{equation}
\begin{equation}
W_{orc}^{out} = W_{orc,\exp }^{out}
\end{equation}
\begin{equation}
EX_{orc}^{in} = W_{orc,eva}^{in}
\end{equation}
\begin{equation}
EX_{orc}^{out} = EX_{orc,\exp }^{out}
\end{equation}
\begin{figure}[ht]
\centering
\includegraphics[width=0.3\textwidth]{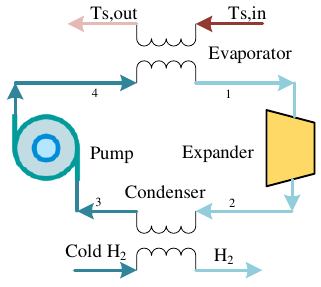}
\setlength{\abovecaptionskip}{-0.1cm}  
\setlength{\belowcaptionskip}{-0.1cm} 
\caption{The schematic diagrams of the ORC system}
\captionsetup{justification=centering}
\vspace{-0.6cm}
\end{figure}\\
6) Plasma \cite{38}\par
The energy input of plasma heating is electricity, and the energy output is thermal energy, as detailed below:
\begin{equation}
W_{pla}^{in} = P_{pla} \cdot \Delta t
\end{equation}
\begin{equation}
W_{pla}^{out} = W_{pla}^{in} \cdot {\eta _{pla}}
\end{equation}
\begin{equation}
Ex_{pla}^{in} = (1 - {{\mathop{\rm Re}\nolimits} _1}) \cdot W_{pla}^{out}
\end{equation}
\begin{align}
Ex_{pla}^{out} = \frac{\gamma }{{\gamma  - 1}}{n_{\exp }}R((T_{pla,\ce{H2}}^{out} - T_{pla,\ce{H2}}^{in}) - \nonumber\\
&\hspace{-2.3cm} {T_0}\ln (T_{pla,\ce{H2}}^{out}/T_{pla,\ce{H2}}^{in}))
\end{align}
7) Shaft furnace \cite{39} \par
The energy sources of the hydrogen-based shaft furnace \cite{40} include the material gas and pellet ore. The energy outputs consist of HDRI, recycled hydrogen, and water vapor.
\begin{equation}
W_{sf}^{in} = W_{sf,ore}^{in} + W_{sf,{H_2}}^{in}
\end{equation}
\vspace{-0.45cm}
\begin{subequations}
    \begin{align}
    &W_{sf,ore}^{in} = W_{sf,ore}^{ph} + W_{sf,ore}^{ch}\\
    &W_{sf,ore}^{ph} = {m_{ore}}{c_{ore,p}}({T_{ore}} - {T_0}) \\
    &W_{sf,ore}^{ch} = {n_{ore}}(\Delta {G_{F{e_2}{O_3}}} + 2\Delta {G_{Fe}} + 3\Delta {G_O})
    \end{align}
\end{subequations}
\vspace{-0.45cm}
\begin{subequations}
    \begin{align}
    &W_{sf,{H_2}}^{in} = W_{sf,{H_2}}^{ph} + W_{sf,{H_2}}^{ch}\\
    &W_{sf,{H_2}}^{ph} = \frac{\gamma }{{\gamma  - 1}}({n_1} + {n_2})R(T_{sf}^{in} - {T_0})\\
    &W_{sf,{H_2}}^{ch} = ({n_1} + {n_2}) \cdot LH{V_{{H_2}}}
    \end{align} 
\end{subequations}
\vspace{-0.45cm}
\begin{equation}
W_{sf}^{out} = W_{DRI}^{out} + W_{C{H_2}}^{out} + W_{{H_2}O}^{out}
\end{equation}
\vspace{-0.45cm}
\begin{subequations}
\begin{align}
    &W_{DRI}^{out} = W_{DRI}^{ph} + W_{DRI}^{ch}\\
    &W_{DRI}^{ph} = {m_{DRI}}{c_{DRI}}({T_{DRI}} - {T_0})\\
    &W_{DRI}^{ch} = {n_{DRI}}\Delta {G_{Fe}}
    \end{align} 
\end{subequations}
\vspace{-0.45cm}
\begin{subequations}
    \begin{align}
    &W_{C{H_2}}^{out} = W_{C{H_2}}^{ph} + W_{C{H_2}}^{ch}\\
    &W_{C{H_2}}^{ph} = \frac{\gamma }{{\gamma  - 1}}{n_2}R(T_{sf}^{out} - {T_0})\\
    &W_{C{H_2}}^{ch} = {n_2} \cdot LH{V_{{H_2}}}
    \end{align}
\end{subequations}
\vspace{-0.45cm}
\begin{subequations}
    \begin{align}
    &W_{{H_2}O}^{out} = W_{{H_2}O}^{ph} + W_{{H_2}O}^{ch}\\
    &W_{{H_2}O}^{ph} = \frac{{{\gamma _{{H_2}O}}}}{{{\gamma _{{H_2}O}} - 1}} \cdot \frac{{{n_1}}}{2} \cdot R(T_{sf}^{out} - {T_0})\\
    &W_{{H_2}O}^{ch} = \frac{{{n_1}}}{2} \cdot (\Delta {G_{{H_2}O}} + 2\Delta {G_H} + \Delta {G_O})
    \end{align} 
\end{subequations}\par
Similarly, the exergy inputs of the hydrogen-based shaft furnace include the material gas and pellet ore, while the exergy outputs consist of sponge iron, recycled hydrogen, and water vapor. The detailed mathematical models are as follows:
\begin{equation}
EX_{sf}^{in} = EX_{sf,ore}^{in} + EX_{sf,{H_2}}^{in}
\end{equation}
\begin{subequations}
    \begin{align}
    &EX_{sf,ore}^{in} = EX_{sf,ore}^{ph} + EX_{sf,ore}^{ch}\\
    &EX_{sf,ore}^{ph} = {m_{ore}}{c_{ore,p}}({T_{ore}} - {T_0} - {T_0}\ln \frac{{{T_{ore}}}}{{{T_0}}})\\
    &EX_{sf,ore}^{ch} = {n_{ore}} \cdot (\Delta {G_{ore}} + \sum\limits_{i \in \{ O,Fe\} } {{\varphi _i}\Delta{G{x_i} + }} \nonumber\\
    &\hspace{3.5cm} R{T_0}\sum\limits_{i \in \{ O,Fe\} } {{\varphi _i}\ln {x_i}} )
    \end{align} 
\end{subequations}
\begin{subequations}
    \begin{align}
    &EX_{sf,{H_2}}^{in} = EX_{sf,{H_2}}^{in,ph} + EX_{sf,{H_2}}^{in,ch}\\
    &EX_{sf,{H_2}}^{in,ph} = \frac{\gamma }{{\gamma  - 1}}({n_1} + {n_2})R(T_{sf}^{in} - {T_0} - {T_0}\ln \frac{{T_{sf}^{in}}}{{{T_0}}})\\
    &EX_{sf,{H_2}}^{in,ch} = LH{V_{{H_2}}} \cdot ({n_1} + {n_2})
    \end{align}
\end{subequations}
\begin{equation}
EX_{sf}^{out} = EX_{sf,{H_2}}^{out} + EX_{sf,DRI}^{out} + EX_{sf,{H_2}O}^{out}
\end{equation}
\begin{subequations}
    \begin{align}
    &EX_{sf,{H_2}}^{out} = EX_{sf,{H_2}}^{out,ph} + EX_{sf,{H_2}}^{out,ch}\\
    &EX_{sf,{H_2}}^{out,ph} = \frac{\gamma }{{\gamma  - 1}}{n_2}R(T_{sf}^{out} - {T_0} - {T_0}\ln \frac{{T_{sf}^{out}}}{{{T_0}}})\\
    &EX_{sf,{H_2}}^{out,ch} = LH{V_{{H_2}}} \cdot {n_2}
    \end{align} 
\end{subequations}
\begin{subequations}
    \begin{align}
    &\hspace{-1.0cm} EX_{sf,{H_2}O}^{out} = EX_{sf,{H_2}O}^{ph} + EX_{sf,{H_2}O}^{ch}\\
    &\hspace{-1.0cm} EX_{sf,{H_2}O}^{ph} = \frac{{{\gamma _{{H_2}O}}}}{{{\gamma _{{H_2}O}} - 1}} \cdot \frac{{{n_1}}}{2} \cdot R(T_{sf}^{out} - {T_0} - {T_0}\ln \frac{{T_{sf}^{out}}}{{{T_0}}})\\
    &\hspace{-1.0cm}EX_{sf,{H_2}O}^{ch} = \frac{{{n_1}}}{2}(\Delta {G_{{H_2}O}} + \sum\limits_{i \in \{ O,H\} } {{\varphi _i}\Delta{G{x_i}}} \nonumber \\
    &\hspace{3.0cm} + R{T_0}\sum\limits_{i \in \{ O,H\} } {{\varphi _i}\ln {x_i}} )
    \end{align}
\end{subequations}
\vspace{-0.45cm}
\subsection{Efficiency indicators}
\noindent 1) Energy efficiency \par
The entire zero-carbon hydrogen metallurgy production system includes the hydrogen production, storage, heating, and recovery phases. Throughout this process, hydrogen flows through various stages, involving the interconversion among electricity, heat, and cold energy. Electricity serves as both the starting and endpoint of energy conversion within the proposed system, providing a unified economic value to the energy flowing through the system. The energy inputs of the proposed system include the electricity energy used by the electrolyzer, plasma and compressor, as well as the energy carried by pellet ore. The energy outputs consist of the heat energy from the high-temperature thermal storage tank, the chemical energy of the furnace top gas, electricity generated by the ORC and expander, the energy of the DRI, the chemical energy in the water vapor, as detailed below:
\begin{equation}
W_{vsys}^{in} = W_{el}^{in} + W_{comp}^{in} + W_{pla}^{in} + W_{sf,ore}^{in}
\end{equation}
\begin{align}
\hspace{-1cm}W_{vsys}^{out} = W_{D,TS}^{out} + W_{\exp }^{out} + W_{orc}^{out} + W_{DRI}^{out} + W_{C{H_2}}^{ch} + W_{{H_2}O}^{ch}
\end{align}
\begin{equation}
{\eta _{ven}} = \frac{{W_{vsys}^{out}}}{{W_{vsys}^{in}}} 
\end{equation} \par
The utilization ratio of material hydrogen is:
\begin{equation}
{\eta _{H_{2}}} = \frac{{{n_1}}}{{{n_1} + {n_2}}}
\end{equation} \par
Since the circulating hydrogen does not participate in chemical reactions and only serves as a heat carrier for the shaft furnace, it is continuously recycled within the proposed system. To accurately reflect the actual energy efficiency from a full life-cycle of energy, only the electricity used to produce the reduction hydrogen should be considered as an energy input, while the chemical energy of the circulating hydrogen is excluded from the energy output. The actual energy efficiency is as follows:
\begin{equation}
W_{sys}^{out} = W_{D,TS}^{out} + W_{\exp }^{out} + W_{orc}^{out} + W_{DRI}^{out} + W_{{H_2}O}^{ch}
\end{equation}
\begin{equation}
W_{sys}^{in} = W_{el}^{in}{\eta _{gas}} + W_{comp}^{in} + W_{pla}^{in} + W_{sf,ore}^{in}
\end{equation}
\begin{equation}
EE= \frac{{W_{sys}^{out}}}{{W_{sys}^{in}}}
\end{equation}
2) Exergy efficiency \par
Given that the proposed system includes low-grade and high-grade heat, with the shaft furnace primarily utilizing high-grade heat, this paper introduces  an exergy efficiency index based on the exergy models of system components to better evaluate energy quality.
\begin{align}
EX_{sys}^{out} = EX_{D,TS}^{out} + EX_{\exp }^{out} + EX_{orc}^{out} \nonumber \\
&\hspace{-1.6cm}+ EX_{DRI}^{out} + EX_{{H_2}O}^{ch}
\end{align}
\begin{equation}
EX_{sys}^{in} = EX_{el}^{in}{\eta _{gas}} + EX_{comp}^{in} + EX_{pla}^{in} + EX_{sf,ore}^{in}
\end{equation}
\begin{equation}
EX = \frac{{EX_{sys}^{out}}}{{EX_{sys}^{in}}}
\end{equation}
3) Energy-carbon efficiency \par
Based on the energy efficiency mentioned above and the carbon efficiency \cite{41}, in accordance with the benchmark and baseline energy efficiency suggested by China’s National Development and Reform Commission for industrial sectors, this paper establishes an integrated energy-carbon efficiency index,incorporating energy equivalent of carbon emissions cost. The coupling of energy efficiency and carbon efficiency benefits from the fact that electricity serves as both the starting and endpoint of energy conversion within the proposed system, providing a unified economic value. The details are as follows:
\begin{equation}
{E_{C}} = \frac{{W_{sys}^{out} - {\varphi _{CE}}CE}}{{W_{sys}^{in}}}
\end{equation}
\begin{equation}
{\varphi _{CE}} = \frac{{{M_{CO2}}}}{{{M_{energy}}}}
\end{equation}
\begin{equation}
{M_{CO2}} = {C_{base}}{P_{CO2}}
\end{equation}
\begin{equation}
{M_{energy}} = {{{M_{c - p}}} \mathord{\left/
 {\vphantom {{{M_{c - p}}} {{Q_c}}}} \right.
 \kern-\nulldelimiterspace} {{Q_c}}}
\end{equation}
\begin{equation}
CE = \left\{
\begin{aligned}
&\frac{{{C_{DRI}}}}{{{C_{base}}}},CE \le 1\\
&\frac{{{C_{DRI}}}}{{{C_{base}}}} \times \vartheta ,1 \le CE \le 1.2\\
&\frac{{{C_{DRI}}}}{{{C_{base}}}} \times \nu ,1.2 \le CE
\end{aligned}
\right.
\end{equation}
\begin{table*}[t!]
\renewcommand{\arraystretch}{1.5}
\begin{center}
\fontfamily{ptm}\selectfont
\footnotesize
\caption{\centering The technical indexes values of the proposed zero-carbon hydrogen metallurgy system}
\begin{tabular}{llll}
\toprule
Temperature(K) & 298 & Pressure of shaft furnace(bar) & 8\\
\midrule 
\parbox{5cm}{Compression ratios of compressor for circle \ce{H2}}\vspace{5pt} & \parbox{3cm}{3.4/3.4/3.4} & Temperature of furnace top gas(K) & 723\\
\parbox{5cm}{Compression ratios of compressor for reduction \ce{H2}}\vspace{5pt}& 2.56/2.56/2.56 & Heat loss ratio of the shaft furnace &0.15 \\
\parbox{5cm}{Isentropic efficiencies of compressor for circle \ce{H2}}\vspace{5pt} &0.918/0.918/0.918 &Heat loss ratio of furnace dust & 0.03 \\
\parbox{5cm}{Isentropic efficiencies of compressor for reduction \ce{H2}}\vspace{5pt} & 0.915/0.915/0.915 & Pressure of hydrogen storage tank(MPa) &20\\
Expansion ratios of expander &2.85/2.85/3 &Heat efficiency of plasma & 0.95\\
Hydrogen storage temperature (K) & 298 & Recovery efficiency of heat storage tank & 0.85\\
Price of coke oven gas (direct procurement) & 2.5 (CNY/Nm3) & Carbon trading price \cite{Carbon_price}  & 120 (CNY/t) \\
\bottomrule
\end{tabular}
\end{center}
\end{table*}
\section{Case study}
\label{sec4}
The main parameters of the proposed zero-carbon hydrogen metallurgy system are presented in Table 1. The system's DRI output is set at 1000kg. To prevent pellet ore from sticking or melting, the hydrogen temperature of entering shaft furnace is controlled within the range of 750°C to 1000°C.\par
\subsection{ Testing and comparison with different systems and reduction gases}
\label{sec4.1}
The energy efficiency, exergy efficiency, and energy-carbon efficiency of the proposed zero-carbon hydrogen metallurgy system at various reduction gas temperatures are presented  in Fig. 5. The energy efficiency, exergy efficiency, energy-carbon efficiency gradually decrease as the temperature rises, reaching their lowest values of 0.3975, 0.2772, and 0.3975, respectively, at 1000°C. This decline occurs because higher temperature reduces the volume of material gas entering the shaft furnace, which in turn decreases heat generated by the compressor and electricity generated by the expander. Furthermore, these reduced energy forms a certain proportion of the system’s total energy inputs and outputs, amplifying their impact and decreasing the system's overall performance. Since the proposed system produces zero-carbon emissions, its carbon emissions cost is zero, leading to identical energy efficiency and  energy-carbon efficiency under all operating conditions. In the traditional DRI production system , mixed reduction gas serves as the shaft furnace's material gas, and employs a combustion furnace fueled by coke oven gas to heat the material gas, along with heat exchangers to recover heat from the furnace top gas. The energy efficiency, exergy efficiency, and energy-carbon efficiency of the traditional DRI production system, which uses \ce{H2}/CO gas mixtures with ratios of 6/4 and 8/2 are shown in Fig. 6 and Fig. 7. The results show that increasing the material gas temperature progressively enhances the energy efficiency, exergy efficiency, and energy-carbon efficiency of the traditional DRI production system, while simultaneously reducing the material gas volume. This improvement is attributed to the higher temperature reduction gas, which enhances the heat entering the shaft furnace, increases heat recovery from furnace top gas, and reduces the circulating gas volume required to maintain the furnace's internal heat balance. Furthermore, a comparison of material gas volumes in Fig. 5, Fig. 6, Fig. 7 shows that the a higher CO proportion in the material gas leads to a lower material gas volume. This is due to the exothermic reaction between the CO and pellet ore reduces the amount of gas required to sustain the shaft furnace's heat balance. Additionally, at 1000°C, the material gas volume of the traditional DRI production system with an \ce{H2}/CO ratio of 6/4 is higher than that of the aforementioned researches. This difference is primarily attributed to the furnace top gas temperature and the DRI temperature, which are determined by the gas-solid heat transfer differential equations or tranted as constants.  
\begin{figure}[ht]
\includegraphics[width=0.45\textwidth]{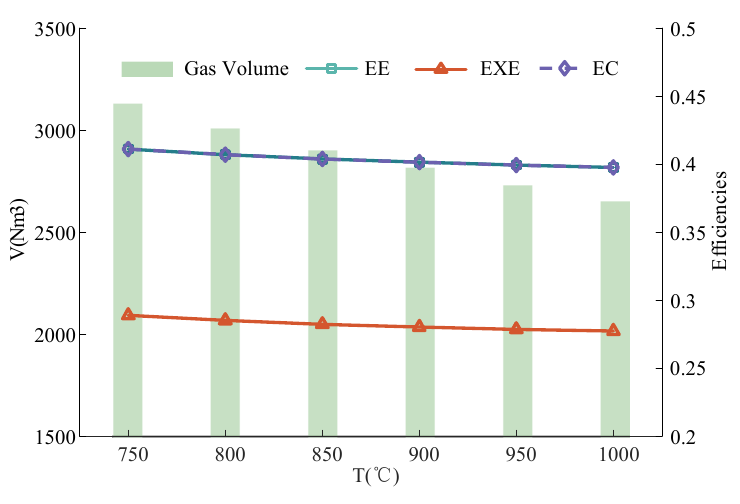}
\setlength{\abovecaptionskip}{-0.1cm}  
\setlength{\belowcaptionskip}{-0.1cm} 
\caption{The efficiencies of the zero-carbon hydrogen metallurgy system}
\captionsetup{justification=centering}
\vspace{-0.6cm}
\end{figure}
\begin{figure}[ht]
\includegraphics[width=0.45\textwidth]{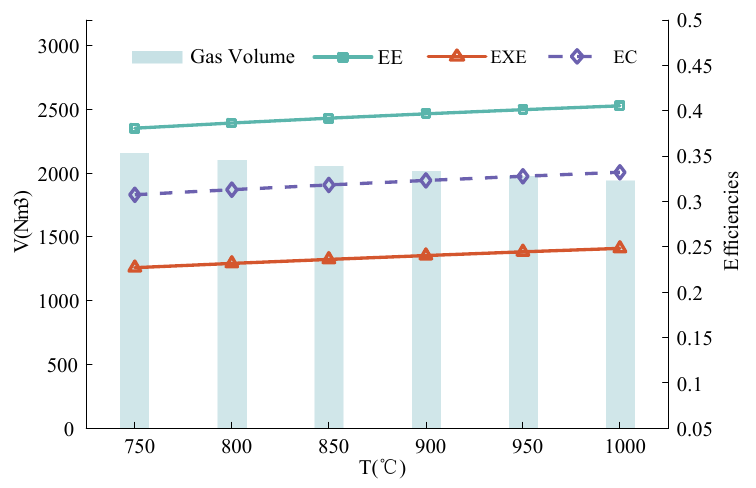}
\setlength{\abovecaptionskip}{-0.1cm}  
\setlength{\belowcaptionskip}{-0.1cm} 
\caption{The efficiencies of the traditional DRI production system with an  $ \mathrm{H}_{2}/\mathrm{CO} $ ratio of 6/4}
\captionsetup{justification=centering}
\vspace{-0.6cm}
\end{figure}
\begin{figure}[ht]
\includegraphics[width=0.45\textwidth]{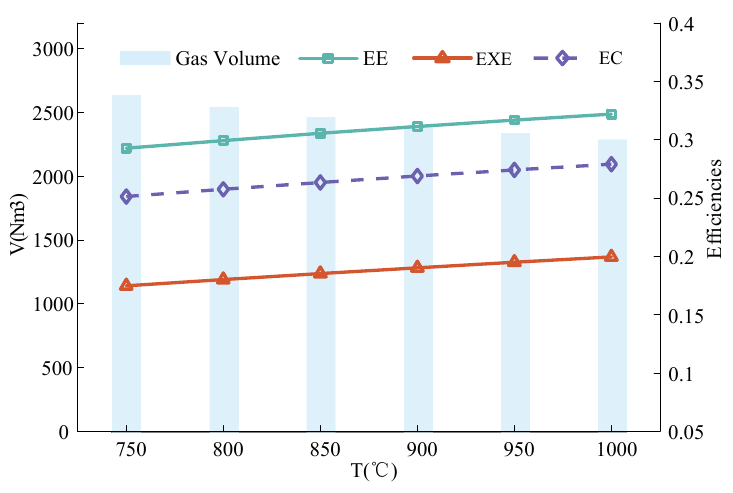}
\setlength{\abovecaptionskip}{-0.1cm}  
\setlength{\belowcaptionskip}{-0.1cm} 
\caption{The efficiencies of the traditional DRI production system with an  $ \mathrm{H}_{2}/\mathrm{CO} $ ratio of 8/2}
\captionsetup{justification=centering}
\vspace{-0.4cm}
\end{figure}\par
As shown in Fig. 8, the energy efficiency of the proposed zero-carbon hydrogen metallurgy system is higher than that of the traditional DRI production system with an $\mathrm{H}_{2}/\mathrm{CO}$ ratio of 8/2, along with increasing trend of reduction gas temperature. However, when reduction gas temperature exceeds ${950}^{\circ}\text{C}$, the energy efficiency of the traditional DRI production system with an $\mathrm{H}_{2}/\mathrm{CO}$ ratio of 6/4 outperforms the proposed zero-carbon hydrogen metallurgy system. This outcome is primarily because, in the proposed system, the electricity generated by the ORC and expander constitutes a significant proportion of the total energy output. As the temperature increases, the energy output decreases more greatly than the energy input, leading to a downward trend in energy efficiency. Conversely, in the traditional DRI production system with \ce{H2}/CO of 6/4, the substantial heat generated by the exothermic reaction between CO and pellet ore remains stable, and the increased heat recovery from the furnace top gas contributes to a rise in energy efficiency. At 1000°C, their energy efficiencies are 0.3975 and 0.4053 , respectively. 
Fig. 9 compares the exergy efficiency of various reduction gases. At 1000°C, the exergy efficiencies of 0.2772, 0.2483, 0.1996 demonstrate that the zero-carbon hydrogen metallurgy system outperforms the traditional DRI production system with $\mathrm{H}_{2}/\mathrm{CO}$ ratios of 6/4 and 8/2. It is opposite to the aforementioned energy efficiency, and this advantage arises from the fact that the proposed zero-carbon hydrogen metallurgy system is powered by electricity, a high-grade energy source. Fig. 10 shows the comparative results of energy-carbon efficiency of various material gases. The energy-carbon efficiency of the proposed zero-carbon hydrogen metallurgy system surpasses that of the mentioned traditional DRI production system, attributing to its zero carbon emissions and the impact of carbon trading price in Shanghai Carbon Emissions Exchange. As the carbon trading price increases and distinction between green steel and carbon-based steel will become clearer in the future, the proposed zero-carbon hydrogen metallurgy system will present significant advantages in both energy-carbon efficiency and economic profitability.  
\begin{figure}[ht]
\includegraphics[width=0.45\textwidth]{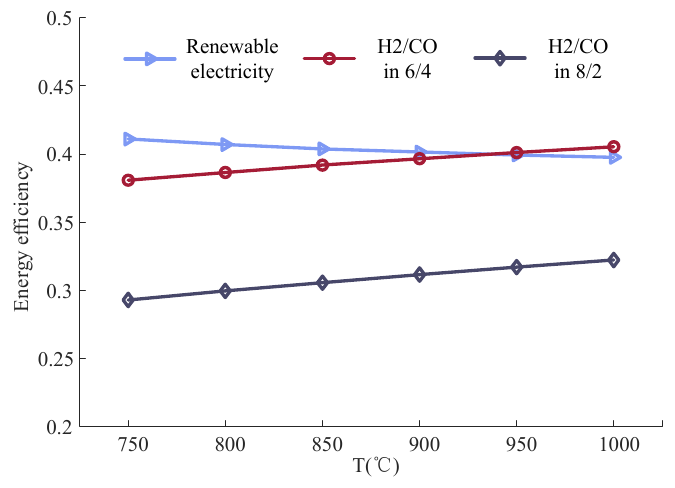}
\setlength{\abovecaptionskip}{-0.1cm}  
\setlength{\belowcaptionskip}{-0.1cm} 
\caption{The comparisons of the energy efficiency of various material gases}
\captionsetup{justification=centering}
\vspace{-0.6cm}
\end{figure}
\begin{figure}[ht]
\includegraphics[width=0.45\textwidth]{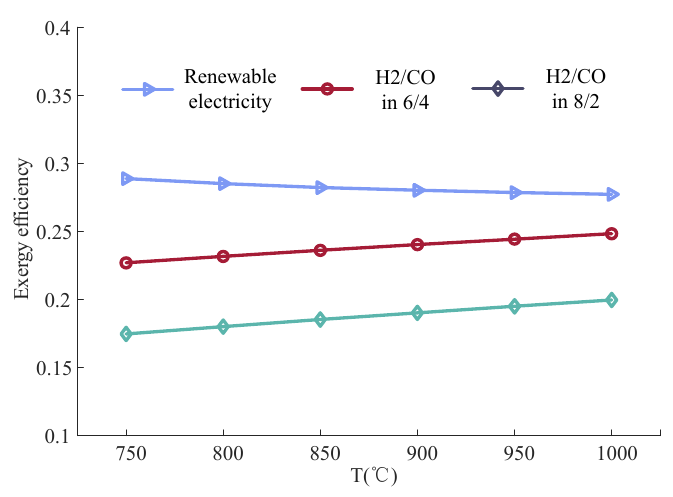}
\setlength{\abovecaptionskip}{-0.1cm}  
\setlength{\belowcaptionskip}{-0.1cm} 
\caption{The comparisons of the exergy efficiency of various material gases}
\captionsetup{justification=centering}
\vspace{-0.4cm}
\end{figure}\par
\noindent Fig. 11 and Fig. 12 respectively present the heat released by the CO reduction reaction and combustion in the traditional DRI production system with \ce{H2}/CO ratios of 6/4 and 8/2, as well as the energy equivalent of the carbon emissions cost. At 1000°C, in the traditional DRI production system with \ce{H2}/CO ratio of 6/4, the heat released by the CO reduction reaction and the combustion of CO from coke oven gas  is $2.87\times10^{8}$J, while the energy equivalent of the carbon emissions cost is $4.83\times10^{8}$J. Similarly, in the traditional DRI production system with \ce{H2}/CO ratio of 8/2, the heat released by the CO reduction and the combustion of CO from coke oven gas is $2.69\times10^{8}$J, and the energy equivalent of the carbon emissions cost is $3.08\times10^{8}$J.
It is apparent that the heat released by the CO reactions is lower than the energy equivalent of the carbon emissions cost. This finding confirms that relying on the heat released from carbon-based energy is not cost-effective when carbon emissions cost is considered. With the continuous expansion of China’s carbon market and the participation of more industries, carbon trading prices are anticipated to rise steadily. This upward trend  will further constrain the utilization of carbon-based energy in the tradition DRI production system, aligning with corporate strategies aimed at reducing costs and increasing profitability. It is clear that the proposed zero-carbon hydrogen metallurgy system is likely to gain greater popularity.  
\begin{figure}[ht]
\includegraphics[width=0.45\textwidth]{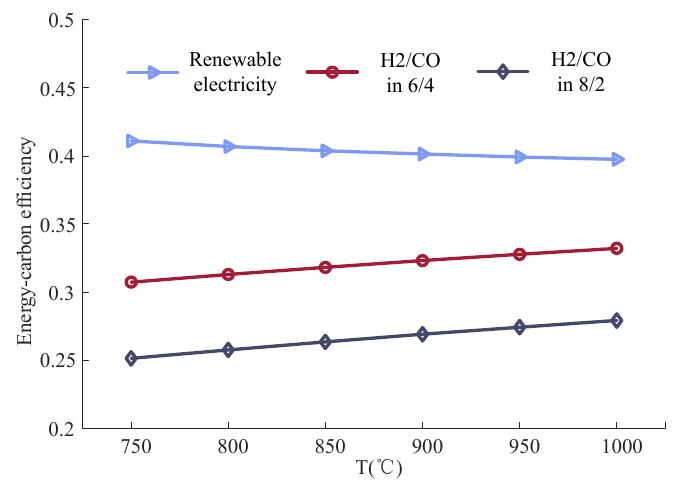}
\setlength{\abovecaptionskip}{-0.1cm}  
\setlength{\belowcaptionskip}{-0.1cm} 
\caption{The comparisons of the energy-carbon efficiency of various material gases}
\captionsetup{justification=centering}
\vspace{-0.6cm}
\end{figure}
\begin{figure}[ht]
\includegraphics[width=0.45\textwidth]{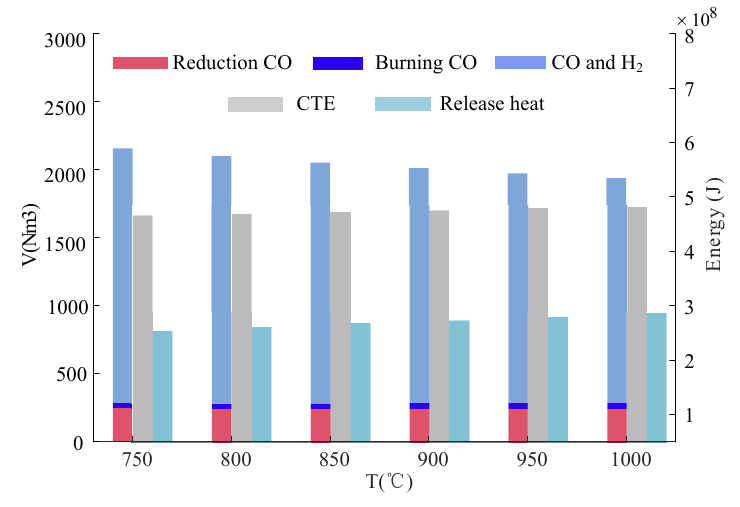}
\setlength{\abovecaptionskip}{-0.1cm}  
\setlength{\belowcaptionskip}{-0.1cm} 
\caption{The heat released by CO compared to the energy equivalent of carbon emissions cost of the traditional DRI production system with an  $ \mathrm{H}_{2}/\mathrm{CO} $  ratio of 6/4}
\captionsetup{justification=centering}
\vspace{-0.4cm}
\end{figure} \par
\subsection{Sensitivity analysis}
Fig. 13 shows the energy efficiencies of the proposed zero-carbon system with and without waste heat recovery. At 1000°C, the energy efficiencies of the system are 0.3975 and 0.21, respectively, for cases with and without waste heat recovery. This demonstrates that incorporating waste heat recovery significantly enhances energy efficiency, and it highlights essential role and necessity of waste heat recovery within the hydrogen metallurgy production system. Furthermore, Fig. 14 presents a sensitivity analysis of the newly added the ORC and expander equipment in the proposed zero-carbon hydrogen metallurgy system. As reduction gas temperature increases, the energy output (excluding the ORC and expander) increases, while the energy output of the ORC and expander decreases. This upward trend is attributed to the high-temperature furnace top gas and HDRI. The downward trend dues to the decreasing volume of reduction gas, which results in less waste heat generated by compressor and internal energy from the high-pressure hydrogen, thereby reducing energy output of the proposed system. Meanwhile, the ORC generates more electricity than the expander, primarily due to its low-temperature liquid working fluid absorbs significant thermal energy from the ambient air as well as from the low-temperature thermal storage tank.
The conventional view holds that the energy recovery from furnace top gas significantly influences energy efficiency. Thus,this paper compares the energy recovered from furnace top gas with the electricity energy generated by the newly added the ORC and expander, as shown in Fig. 15. At 1000°C, the electricity generated by the ORC and expander is $1.74\times 10^{9}$J, and the heat recovered from furnace top gas is $1.11\times 10^{9}$J. It highlights the substantial energy potential in the low-temperature waste heat and cold energy of the hydrogen metallurgy system. This innovative measure transcends traditional energy efficiency improvements that primarily rely on heat recovery from furnace top gas, providing clear guidance for future strategies to enhance energy efficiency in hydrogen metallurgy production system. Additionally, the ORC unit converts low-temperature waste heat into electricity, assigning an economic value to waste heat through its correlation with electricity prices, and ORC reinforces the validity of the energy-carbon coupling factor used in the proposed energy-carbon efficiency indicator, offering a robust quantitative measure of the system’s overall energy.
\begin{figure}[ht]
\includegraphics[width=0.45\textwidth]{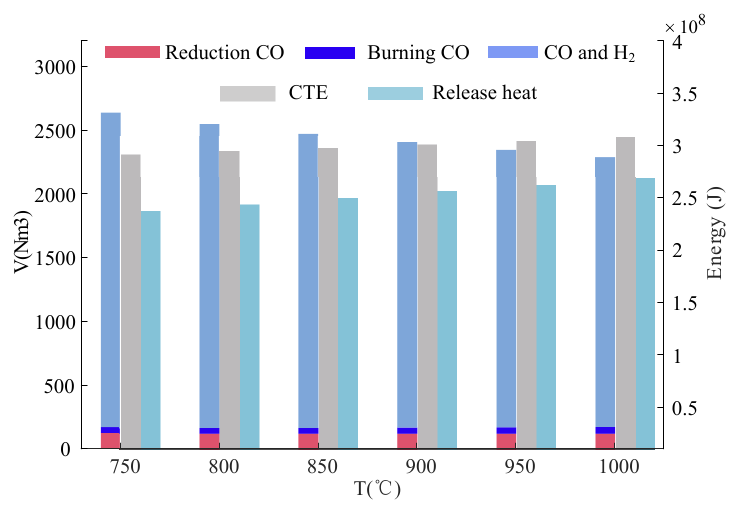}
\setlength{\abovecaptionskip}{-0.1cm}  
\setlength{\belowcaptionskip}{-0.1cm} 
\caption{The heat released by CO compared to the energy equivalent of carbon emissions cost of the traditional DRI production system with an  $ \mathrm{H}_{2}/\mathrm{CO} $  ratio of 8/2}
\captionsetup{justification=centering}
\vspace{-0.4cm}
\end{figure}
\begin{figure}[ht]
\includegraphics[width=0.45\textwidth]{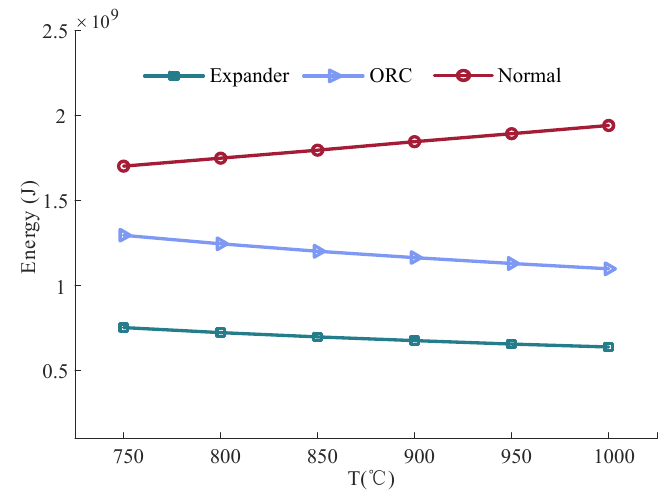}
\setlength{\abovecaptionskip}{-0.1cm}  
\setlength{\belowcaptionskip}{-0.1cm} 
\caption{The comparison results between the energy efficiency with the waste heat recovery and the energy efficiency without the waste heat recovery in the proposed system}
\captionsetup{justification=centering}
\vspace{-0.4cm}
\end{figure}
\begin{figure}[ht]
\includegraphics[width=0.45\textwidth]{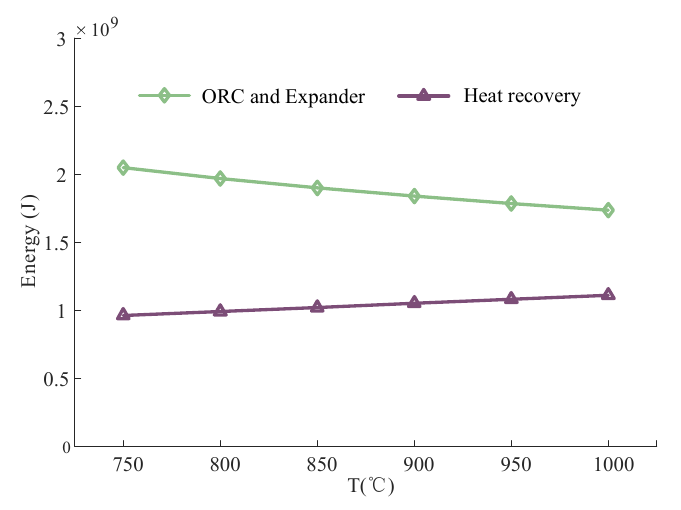}
\setlength{\abovecaptionskip}{-0.1cm}  
\setlength{\belowcaptionskip}{-0.1cm} 
\caption{The Sensitivity analysis of the ORC and the expander equipment in the proposed system}
\captionsetup{justification=centering}
\vspace{-0.4cm}
\end{figure}
\begin{figure}[ht]
\includegraphics[width=0.45\textwidth]{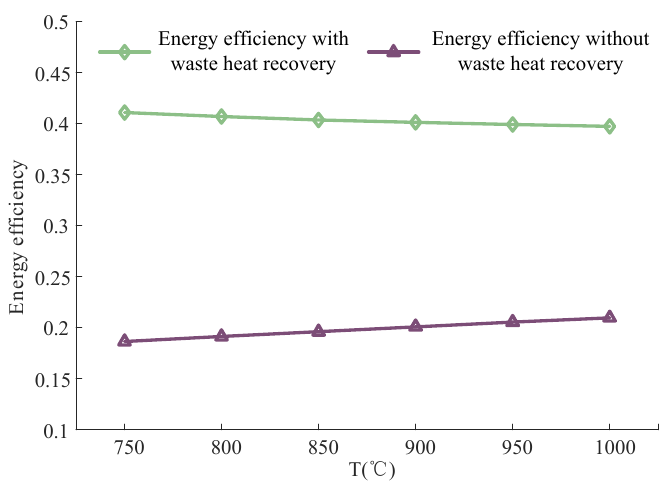}
\setlength{\abovecaptionskip}{-0.1cm}  
\setlength{\belowcaptionskip}{-0.1cm} 
\caption{The comparison results between the energy generated by the ORC and expander equipment and the energy recovery from the furnace top gas in the proposed system}
\captionsetup{justification=centering}
\vspace{-0.4cm}
\end{figure}
\subsection{Influence of grid electricity}
With a carbon emissions factor of 0.57tons/MWh for grid electricity, Fig. 16 shows that the energy efficiency, exergy efficiency, and energy-carbon efficiency of the proposed hydrogen metallurgy system when powered by grid electricity. The results indicate that the energy efficiency and exergy efficiency of the proposed system are both higher than its energy-carbon efficiency at equal temperature. Furthermore,  the findings indicate that relying on limited or no less green electricity as the energy input keeps the system’s energy-carbon efficiency relatively low, thereby increasing the pressure of indirect carbon emissions. It witnesses the importance of directly supplying green electricity from renewable energy station to the iron and steel industry as an effective strategy for significantly reducing carbon emissions.  
\begin{figure}[ht]
\includegraphics[width=0.45\textwidth]{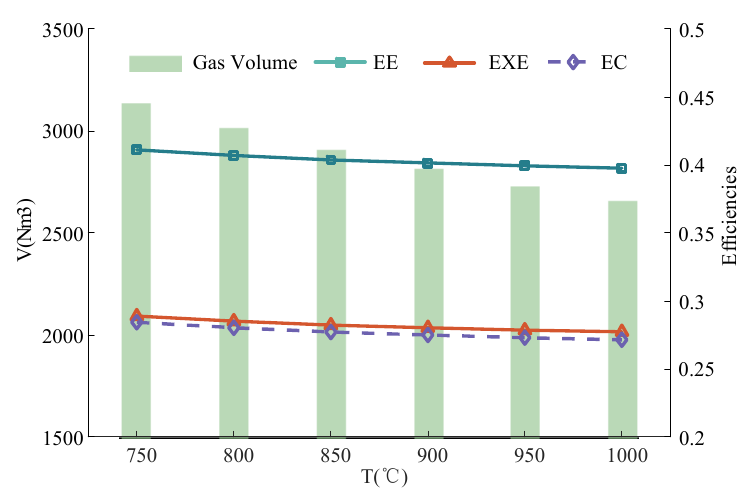}
\setlength{\abovecaptionskip}{-0.1cm}  
\setlength{\belowcaptionskip}{-0.1cm} 
\caption{The energy efficiency, exergy efficiency, and energy-carbon efficiency of the proposed zero-carbon hydrogen metallurgy system powered by the grid electricity}
\captionsetup{justification=centering}
\vspace{-0.4cm}
\end{figure}
\begin{figure}[ht]
\includegraphics[width=0.45\textwidth]{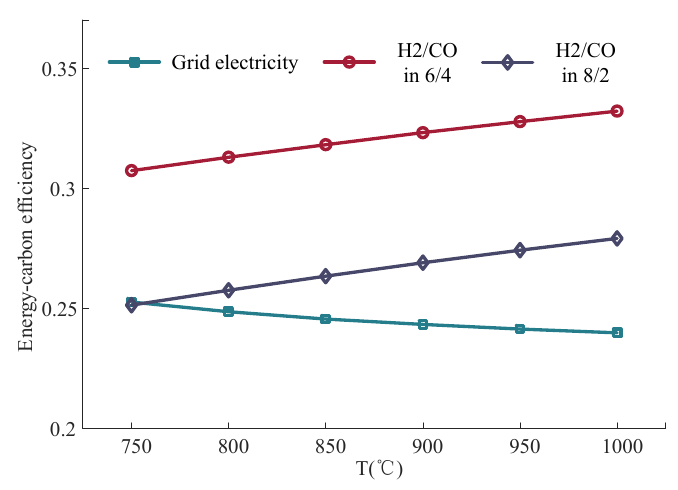}
\setlength{\abovecaptionskip}{-0.1cm}  
\setlength{\belowcaptionskip}{-0.1cm} 
\caption{The comparisons of the energy-carbon efficiency of various energy inputs}
\captionsetup{justification=centering}
\vspace{-0.4cm}
\end{figure}
\begin{figure}[ht]
\includegraphics[width=0.45\textwidth]{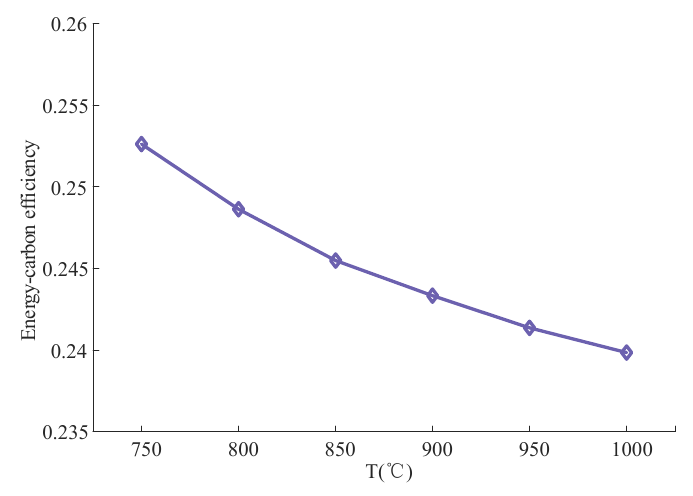}
\setlength{\abovecaptionskip}{-0.1cm}  
\setlength{\belowcaptionskip}{-0.1cm} 
\caption{The energy-carbon efficiency punished by the carbon emissions factor}
\captionsetup{justification=centering}
\vspace{-0.4cm}
\end{figure}
The energy-carbon efficiency of the proposed system using grid electricity is compared with traditional DRI production system systems using \ce{H2}/CO ratios of 6/4 and 8/2, as shown in Fig. 17. The results indicate that the energy-carbon efficiency of the proposed system is lower than that of the traditional DRI production system with $\mathrm{H}_{2}/\mathrm{CO}$ ratios of 6/4 and 8/2. When grid electricity is used to power the proposed system, its carbon emissions exceed the baseline published by the National Development and Reform Commission and its CE value is above 1.2, resulting in a penalty using Equation 96. Fig. 18 shows the energy-carbon efficiency after the penalty, highlighting the request for mandatory measures to increase green electricity consumption. The discrepancy between achieving zero carbon emissions in production and paying for carbon-intensive electricity presents challenge for carbon market regulations when setting carbon emissions accounting standards. Furthermore, the proposed zero-carbon hydrogen metallurgy system aligns well with China’s rapid renewable energy development. It not only effectively utilizes surplus renewable energy  but also enables substantial carbon reduction in the iron and steel industry, playing a key role in developing the new type power system and achieving dual-carbon goals. 
\section{Conclusion}
\label{sec5}
The study analyzes the energy and exergy inputs and outputs of each component, defining energy efficiency, exergy efficiency, and energy-carbon efficiency from an full life-cycle  perspective of energy. The main conclusions are as follows:\par
1) The proposed system exhibits higher exergy efficiency and energy-carbon efficiency compared to the traditional DRI production system with the $\mathrm{H}_{2}/\mathrm{CO}$ ratio of 4/6, but its energy efficiency is lower with reduction gas temperature exceeding  950°C. The traditional DRI production system benefits from the significant heat released by CO exothermic reaction, but the energy equivalent of carbon emissions cost is higher than the energy generated by the exothermic reaction of CO. It reflects that carbon-based energy is not cost-effective when carbon emissions cost are considered.\par
2) At 1000°C, the electricity generated by the ORC and expander is $1.74\times 10^{9}$J, which exceeds that the heat recovered from furnace top gas is $1.11\times 10^{9}$J. This result highlights the superiority of the proposed zero-carbon hydrogen metallurgy system and provides guidance of energy efficiency enhancement for existing DRI production system. \par
3) The proposed energy-carbon efficiency metric directly links system output energy with carbon emissions cost by incorporating a penalty mechanism for carbon-based energy. It not only restricts the heavy reliance on carbon-based energy in the iron and steel industry but also provides a clear guide for smaller industries' investment in efficiency improvements. \par
In the future, the research will focus on the operational strategies of the zero-carbon hydrogen metallurgy system, aiming to address the challenge of balancing stable energy supply with the uncertainties of renewable energy. 
\printcredits \\
\\
\vspace{0.3cm}
\noindent \textbf{Declaration of competing interest}\par
\noindent The authors declare that they have no known competing financial interests or personal relationships that could have appeared to influence the work reported in this paper.
\vspace{0.3cm}

% \noindent \textbf{Data availability}

% The original data can be downloaded from~\cite{Elia-data} and~\cite{North-China-data}. 
% \vspace{0.3cm}

\noindent \textbf{Acknowledgements}

\noindent This work is supported in part by Special Foundation of Xinjiang Key Research and Development Program under Grant (No. 20242126827) and National Natural Science Foundation of China under Grant (No. 52307144).

\bibliographystyle{model1-num-names}
% \bibliographystyle{cas-model2-names}

\bibliography{ECM}

\end{document}